\documentclass[prb,preprint,aps,superscriptaddress,preprintnumbers,amsmath,amssymbn,12pt]{revtex4-2}
\usepackage{amssymb}
\usepackage{amsmath}
\usepackage{graphicx}
\usepackage{multirow}
\usepackage{comment}
\usepackage{longtable}
\usepackage{color}
\usepackage{xfrac}
\usepackage{threeparttable}
\graphicspath{./}
\usepackage[colorlinks,linkcolor=blue,anchorcolor=black,citecolor=blue,urlcolor=black]{hyperref}

\newcommand{\br}{\mathbf{r}}
\newcommand{\brp}{\mathbf{r}^\prime}

\begin{document}
\newcommand{\tny}[1]{\mbox{\tiny $#1$}}
\newcommand{\eqn}[1]{\mbox{Eq.\hspace{1pt}(\ref{#1})}}
\newcommand{\eqs}[2]{\mbox{Eq.\hspace{1pt}(\ref{#1}--\ref{#2})}}
\newcommand{\eqsu}[2]{\mbox{Eqs.\hspace{1pt}(\ref{#1},\ref{#2})}}
\newcommand{\eqtn}[2]{\begin{equation} \label{#1} #2 \end{equation}}
\newcommand{\func}[1]{#1 \left[ \rho \right] }
\newcommand{\mfunc}[2]{#1_{#2} \left[ \rho \right] }
\newcommand{\mmfunc}[3]{#1_{#2} \left[ #3 \right] }
\newcommand{\mmmfunc}[4]{#1_{#2}^{#3} \left[ #4 \right] }
\newcommand{\pot}[1]{v_{\rm #1}}
\newcommand{\spot}[2]{v_{\rm #1}^{#2}}
\newcommand{\code}[1]{\texttt{#1}}
\newcommand{\beq}{\begin{equation}}
\newcommand{\eeq}{\end{equation}}
\newcommand{\bea}{\begin{eqnarray}}
\newcommand{\eea}{\end{eqnarray}}

\def\brpppp{{\mathbf{r}^{\prime\prime\prime\prime}}}
\def\brppp{{\mathbf{r}^{\prime\prime\prime}}}
\def\brpp{{\mathbf{r}^{\prime\prime}}}
\def\brp{{\mathbf{r}^{\prime}}}
\def\bzp{{\mathbf{z}^{\prime}}}
\def\bxp{{\mathbf{x}^{\prime}}}
\def\tp{{{t}^{\prime}}}
\def\tpp{{{t}^{\prime\prime}}}
\def\tppp{{{t}^{\prime\prime\prime}}}

\def\tbr{{\tilde{\mathbf{r}}}}
\def\bk{{\mathbf{k}}}
\def\br{{\mathbf{r}}}
\def\bz{{\mathbf{z}}}
\def\bx{{\mathbf{x}}}
\def\bR{{\mathbf{R}}}
\def\bM{{\mathbf{M}}}
\def\bP{{\mathbf{P}}}
\def\bT{{\mathbf{T}}}
\def\bK{{\mathbf{K}}}
\def\bA{{\mathbf{A}}}
\def\bB{{\mathbf{B}}}
\def\bX{{\mathbf{X}}}
\def\bY{{\mathbf{Y}}}
\def\bP{{\mathbf{P}}}
\def\bI{{\mathbf{I}}}
\def\d{{\mathrm{d}}}
\def\rhor{{\rho({\bf r})}}
\def\rhorp{{\rho({\bf r}^{\prime})}}
\def\rhoi{{\rho_I}}
\def\rhoii{{\rho_{II}}}
\def\rhoj{{\rho_J}}
\def\rhoir{{\rho_I({\bf r})}}
\def\rhoiir{{\rho_{II}({\bf r})}}
\def\rhojr{{\rho_J({\bf r})}}
\def\rhoirp{{\rho_I({\bf r}^{\prime})}}
\def\rhojrp{{\rho_J({\bf r}^{\prime})}}
\def\sumi{{\sum_I^{N_S}}}
\def\sumj{{\sum_J^{N_S}}}
\def\im{{\operatorname{Im}}}

\def\etal{{\it et al.}}
\def\vdw{{van der Waals}}
\def\vw{{von Weizs\"{a}cker}}
\def\qe{{\sc Quantum ESPRESSO}}
\def\se{{Schr\"{o}dinger equation}}
\def\ses{{Schr\"{o}dinger equations}}
\def\bnabla{{\boldsymbol{\nabla}}}
\def\bchi{{\boldsymbol\chi}}
\def\bLambda{{\boldsymbol\Lambda}}
\def\bDelta{{\boldsymbol\Delta}}

\title{Fast and stable tight-binding framework for nonlocal kinetic energy density functional reconstruction in orbital-free density functional calculations}

\author{Yongshuo Chen}
\affiliation{Key Laboratory of Material Simulation Methods \& Software of Ministry of Education, College of Physics, Jilin University, Changchun 130012, China.}
\affiliation{State Key Lab of Superhard Materials, College of Physics, Jilin University, Changchun 130012, China.}

\author{Cheng Ma}
\affiliation{Key Laboratory of Material Simulation Methods \& Software of Ministry of Education, College of Physics, Jilin University, Changchun 130012, China.}
\affiliation{State Key Lab of Superhard Materials, College of Physics, Jilin University, Changchun 130012, China.}

\author{Boning Cui}
\affiliation{Key Laboratory of Material Simulation Methods \& Software of Ministry of Education, College of Physics, Jilin University, Changchun 130012, China.}

\author{Tian Cui}
\affiliation{Institute of High Pressure Physics, School of Physical Science and Technology, Ningbo University, Ningbo, 315211, China}

\author{Wenhui Mi}
\email{mwh@jlu.edu.cn}
\affiliation{Key Laboratory of Material Simulation Methods \& Software of Ministry of Education, College of Physics, Jilin University, Changchun 130012, China.}
\affiliation{State Key Lab of Superhard Materials, College of Physics, Jilin University, Changchun 130012, China.}

\author{Qiang Xu}
\email{xuq@jlu.edu.cn}
\affiliation{Key Laboratory of Material Simulation Methods \& Software of Ministry of Education, College of Physics, Jilin University, Changchun 130012, China.}

\author{Yanchao Wang}
\email{wyc@calypso.cn}
\affiliation{Key Laboratory of Material Simulation Methods \& Software of Ministry of Education, College of Physics, Jilin University, Changchun 130012, China.}
\affiliation{State Key Lab of Superhard Materials, College of Physics, Jilin University, Changchun 130012, China.}

\author{Yanming Ma}
\affiliation{Key Laboratory of Material Simulation Methods \& Software of Ministry of Education, College of Physics, Jilin University, Changchun 130012, China.}
\affiliation{State Key Lab of Superhard Materials, College of Physics, Jilin University, Changchun 130012, China.}

\begin{abstract}
Nonlocal kinetic energy density functionals (KEDFs) with density-dependent kernels are currently the most accurate functionals available for orbital-free density functional theory (OF-DFT) calculations. However, despite advances in numerical techniques and using only (semi)local density-dependent kernels, nonlocal KEDFs still present substantial computational costs in OF-DFT, limiting their application in large-scale material simulations. To address this challenge, we propose an efficient framework for reconstructing nonlocal KEDFs by incorporating the density functional tight-binding approach, in which the energy functionals are simplified through a first-order functional expansion based on the superposition of free-atom electron densities. This strategy allows the computationally expensive nonlocal kinetic energy and potential calculations to be performed only once during the electron density optimization process, significantly reducing computational overhead while maintaining high accuracy. Benchmark tests using advanced nonlocal KEDFs, such as revHC and LDAK-MGPA, on standard structures including Li, Mg, Al, Ga, Si, III-V semiconductors, as well as Mg$_{50}$ and Si$_{50}$ clusters, demonstrate that our method achieves orders-of-magnitude improvements in efficiency, providing a cost-effective balance between accuracy and computational speed. Additionally, the reconstructed functionals exhibit improved numerical stability for both bulk and finite systems, paving the way for developing more sophisticated KEDFs for realistic material simulations using OF-DFT.
\end{abstract}
\maketitle

\section{Introduction}


In the past decades, Kohn-Sham density functional theory (KS-DFT)\cite{hk1964,kohn1965self} has become one of the most widely used and powerful approaches for material simulations, owing to its strong balance between computational accuracy and efficiency, particularly for systems containing hundreds atoms\cite{carloni2002role, burke2012perspective}. However, the noninteracting kinetic energy term in KS-DFT relies on a set of single-particle wavefunctions (or Kohn-Sham orbitals), which are obtained by solving the Kohn-Sham equations as a nonlinear eigenvalue problem. Solving the Kohn-Sham equations incurs a high computational cost, scaling cubically with the number of electrons in the systems, making it prohibitive for large-scale systems in practice\cite{carloni2002role, burke2012perspective}. In contrast, orbital-free density functional theory (OF-DFT)\cite{karasiev2012issues,witt2018orbital,mi2023,xu2024recent} calculates the noninteracting kinetic energy directly from kinetic energy density functionals (KEDFs), which are solely functionals of the electron density. OF-DFT bypasses the need to solve for KS orbitals, resulting in linear scaling with the size of simulating cell and significantly reducing computational overhead. Consequently, OF-DFT enables large-scale first-principles calculations for systems with millions of atoms on a single CPU and hundreds of millions in parallel\cite{shao2018large,hung2009accurate,shao2021dftpy}.

OF-DFT offers a distinct computational advantage by directly using KEDFs to evaluate the noninteracting kinetic energy\cite{witt2018orbital,karasiev2012issues,xu2024recent, mi2023}. However, its accuracy and applicability are largely governed by the approximations in KEDFs, which significantly influence the range of systems that can be simulated. In recent years, advanced KEDFs at various levels (GGA\cite{weizsacker1935,laricchia2011generalized, ou1991approximate,perdew1992generalized,ernzerhof2000role,constantin2009kinetic,vitos2000local,thakkar1992comparison,lembarki1994obtaining}, meta-GGA\cite{luo2018simple,garcia2007kinetic}, and nonlocal\cite{chacon1985nonlocal,wang1992kinetic,smargiassi1994orbital,perrot1994hydrogen,wang1999orbital,huang2010nonlocal,mi2018nonlocal,xu2019nonlocal}) have been proposed to expand the applicability of OF-DFT. Notably, it has been shown that the exact KEDF is inherently nonlocal, exhibiting infinite complexity\cite{yang1987ab}. This complexity can be captured within the kernel of the two-point nonlocal KEDF formalism\cite{mi2023,wang1999orbital}.

The two-point nonlocal functional inherently requires at least double integration. On the one hand, more sophisticated kernels incorporating additional density information are necessary; on the other hand, the computational complexity of these functionals must be carefully addressed. The progression of two-point KEDFs from density-independent kernels\cite{wang1992kinetic,smargiassi1994orbital,perrot1994hydrogen} to slightly density-dependent\cite{wang1999orbital,xu2019nonlocal} and (semi)local-density-dependent kernels\cite{chacon1985nonlocal,shao2021revised,huang2010nonlocal,mi2019orbital,xu2020nonlocal} has expanded OF-DFT's applicability from nearly free-electron systems to semiconductors and more complex systems, such as interfaces, quantum dots and clusters\cite{huang2010nonlocal,shao2021revised, mi2019orbital,xu2020nonlocal}. Although various numerical techniques, such as fast Fourier transform and spline methods\cite{mi2019orbital,huang2010nonlocal}, have been employed to reduce the scaling and computational costs of nonlocal KEDFs, achieving quasi-linear scaling for two-point nonlocal functionals is currently feasible only when considering, at most, local density-dependent kernels.

It is anticipated that increasing the kernel's density dependence could significantly enhance the accuracy of OF-DFT, both in theory and practice. However, KEDFs with (semi)local-density-dependent kernels, such as the Chacón-Alvarellos-Tarazona (CAT)\cite{chacon1985nonlocal}, Huang-Carter (HC)\cite{huang2010nonlocal}, and local-density-approximation-kernel (LDAK)\cite{mi2019orbital,xu2020nonlocal} KEDFs, have seen limited advancement due to their high computational complexity. Specifically, their costs are approximately 100 times greater than those of density-independent KEDFs\cite{constantin2018nonlocal}. Recently, Shao et al. introduced the one-orbital ensemble self-consistent field\cite{shao2021efficient} (OE-SCF) method, an efficient OF-DFT solver that treats the Pauli potential as an external potential. This approach evaluates the nonlocal potential only once per self-consistent iteration step, reducing the frequency of this time-consuming operation to about a dozen times during the density optimization process. While this strategy significantly cuts computational costs for large-scale systems, the nonlocal potential remains the most expensive part of the calculation. Given that most computational time in OF-DFT is spent on nonlocal potential evaluations, developing schemes to reduce the frequency of these costly calculations could further dramatically lower the overall computational cost. This would, in turn, expand the applicability of OF-DFT and facilitate the development of more sophisticated KEDFs.

In this work, we propose a framework inspired by tight-binding (TB) methods\cite{elstner1998self} that significantly reduces the computational cost of OF-DFT with such (semi)local-density-dependent nonlocal KEDFs. This approach approximates the nonlocal KEDF using a first-order functional expansion at a TB reference electron density based on the superposition of free-atom electron densities. This approximation allows the most computationally demanding component—the nonlocal kinetic potential—to be calculated only once in the entire electron density optimization, thereby drastically reducing the computational overhead. This framework has been implemented in the ATLAS code\cite{mi2016atlas,shao2016n}. To benchmark the performance of this method, we applied it to advanced nonlocal KEDFs, namely revised HC (revHC)\cite{shao2021revised} and LDAK-Mi-Genova-Pavanello (LDAK-MGPA)\cite{mi2018nonlocal,xu2020nonlocal}, across a range of bulk and finite systems. As expected, the TB-KEDFs exhibit comparable accuracy to the original KEDFs while substantially reducing the computational cost of OF-DFT. Additionally, the proposed framework demonstrates improved numerical stability compared to conventional OF-DFT calculations using nonlocal KEDFs. We believe that this fast and stable framework paves the way for developing more sophisticated KEDFs for OF-DFT and broadening the scope of OF-DFT for realistic material simulations.

This paper is organized as follows: Section \ref{theoretical} provides the theoretical formalism for the proposed framework to reconstruct nonlocal KEDFs in OF-DFT calculations. The computational methods are outlined in Section \ref{Computaional details}. Section \ref{Results} presents benchmarks evaluating the proposed scheme's computational accuracy, efficiency, and stability. Finally, we conclude with a summary and future outlook in Section \ref{Conclusions}.

\section{Theoretical formalism} \label{theoretical}
The general form of most nonlocal KEDFs can be expressed as:
\beq
\label{eq:1}
T_{s}[\rho]=T_{TF}[\rho] + T_{vW}[\rho] + T_{NL}[\rho], 
\eeq
where $T_{TF}[\rho]=\frac{3}{10}(3\pi^2)^{2/3}\int{\rho^{5/3}(\br)}d^3\br$ and $T_{vW}[\rho]=\frac{1}{8}\int{\frac{|\nabla\rhor|^2}{\rhor}}d^3\br$ are the Thomas-Fermi (TF)\cite{Thomas1927,Fermi1927,fermi1928statistische} and von Weizs\"acker (vW)\cite{weizsacker1935} KEDFs, respectively. The nonlocal part of the KEDF ($T_{NL}$) typically involves a double integral, which can be written as\cite{wang1992kinetic,smargiassi1994orbital,perrot1994hydrogen,wang1999orbital,huang2010nonlocal,xu2019nonlocal,xu2020nonlocal}:
\beq
\label{eq:2}
T_{NL}[\rho]=\int\int{\rho^\alpha(\br)\omega[\rho](\br,\brp)\rho^\beta(\br')d^3\br d^3\brp}, 
\eeq
where $\alpha$ and $\beta$ are the parameters that have specific values for various nonlocal KEDFs\cite{wang1992kinetic,smargiassi1994orbital,perrot1994hydrogen,wang1999orbital,huang2010nonlocal,mi2018nonlocal,mi2019orbital,xu2019nonlocal,xu2020nonlocal}. The term of $\omega[\rho](\br,\brp)$ represents the two-point kernel function, commonly used to capture the nonlocality in KEDFs. As previously mentioned, (semi)local density-dependent kernels provide higher accuracy for modeling systems with highly inhomogeneous electron densities—such as metallic or semiconducting bulk materials, surfaces, and clusters \cite{huang2010nonlocal, shao2021revised, mi2019orbital, xu2020nonlocal}—but their high computational complexity presents significant challenges. To achieve efficient OF-DFT calculations using these functionals, we draw inspiration from the density functional tight-binding\cite{elstner1998self} approach and assume that the electron density of the system can be written as:
\beq
\label{eq:3}
\rho(\br)=\rho_0(\br)+\delta\rho(\br), 
\eeq
where $\rho_0$ is the reference electron density and $\delta\rho$ represents small fluctuations around this reference. With this TB density approximation, we then perform a first-order expansion on the computationally expensive nonlocal part functionals as:
\bea
\label{eq:4}
T_{NL}[\rho_0+\delta\rho] \simeq T_{NL}[\rho_{0}]
+\int V^T_{NL}[\rho_0](\br) \delta\rhor d^3\br, 
\eea
where $V^T_{NL}[\rho_0](\br)\equiv\delta T_{NL}/\delta\rhor|_{\rho_0}$ denotes the nonlocal kinetic potential for the reference electron density. By combining Eqs.~(\ref{eq:1})--(\ref{eq:4}), the expensive nonlocal kinetic energy ($T_{NL}[\rho_0]$) and potential ($V_{NL}^T[\rho_0]$) depend solely on $\rho_0$. They can be calculated during the initialization step through a one-shot calculation, thereby saving considerable time in the electron density optimization process.

Notably, the accuracy of OF-DFT calculations using the TB approximation depends strongly on the choice of nonlocal KEDF and the reference electron density ($\rho_0$). The central idea behind the TB model is that electrons are tightly bound to their respective atoms and are only weakly perturbed by neighboring atoms. As a result, a natural reference electron density is used, defined as the superposition of free-atom electron densities:
\beq
\label{eq:5}
\rho_0(\br)=\sum_{a}\rho_a(\br-\bR_a), 
\eeq
where $\rho_a$ and $\bR_a$ are the $a$th atomic electron density and position, respectively. In principle, the atomic electron density in Eq.~(\ref{eq:5}) can be generated by KS-DFT calculation with nonlocal pseudopotential to reduce errors introduced by local pseudopotential in OF-DFT, which is part of our next step in the ongoing research. However, for the convenience of benchmarking the framework's performance on the reconstruction of KEDFs, the atomic electron density was calculated by KS-DFT with local pseudopotential in this work. The nonlocal functionals of LDAK-MGPA\cite{mi2018nonlocal, xu2020nonlocal} and revHC\cite{shao2021revised}, which demonstrate excellent performance in semiconducting clusters and solids, respectively, were reconstructed in our TB scheme by Eqs.~(\ref{eq:1})--(\ref{eq:5}).

\section{Computational details} \label{Computaional details}
The TB-KEDF framework has been implemented in ATLAS 3.0 \cite{mi2016atlas,shao2016n}. The grid spacing of 0.2~\AA~ was employed for all OF-DFT calculations in ATLAS. The KS-DFT calculations were performed by Quantum Espresso 7.3.1 \cite{quantumEspresso} as the benchmarks. The kinetic energy cutoffs were set to 60 Ry. The $k$-points in Quantum Espresso were generated using the ASE \cite{ase-paper} package with a spacing of 0.02 \AA$^{-1}$ for bulk systems, whereas the $\Gamma$-only $k$-point was applied for all clustered systems. The parameters used in this work were meticulously chosen to ensure energy convergence below 2 meV/atom. Additionally, we ensured that total energies converged to within $10^{-5}$ eV/atom during the density optimization iterations for all calculations.

The bulk-derived local pseudopotentials\cite{huang2008transferable} were employed to model ion-electron interactions across all systems considered. The local density approximation exchange-correlation functional formulated by Perdew and Zunger \cite{Perdew1981PZ} was adopted in all calculations. The atomic electron densities used in TB-KEDFs were determined by solving the single-atom Kohn-Sham equations with the same pseudopotentials and exchange-correlation functionals through an in-house code.

We performed OF-DFT calculations for various crystal and clustered systems as representative examples. For the crystal systems, we carried out bulk property calculations for several crystal phases of five elements (Li, Mg, Al, Ga, and Si), including face-centered cubic (FCC), hexagonal close-packed (HCP), body-centered cubic (BCC), simple cubic (SC), and cubic diamond (CD) structures, as well as hexagonal diamond (HD), complex body-centered cubic (CBCC), $\beta$-tin, and body-centered tetragonal 5 (BCT5) structures\cite{boyer1991new, yin1982theory} for Si, along with nine additional cubic zincblende (ZB) semiconductors. For the clustered systems, we generated 120-random structures of both Mg$_{50}$ and Si$_{50}$ using the CALYPSO\cite{calypso1,calypso2} package to evaluate the predictive ability of the KEDFs for energy ordering.

\section{Results and discussion} \label{Results}

\begin{figure*}
    \centering
    \includegraphics[width=\linewidth]{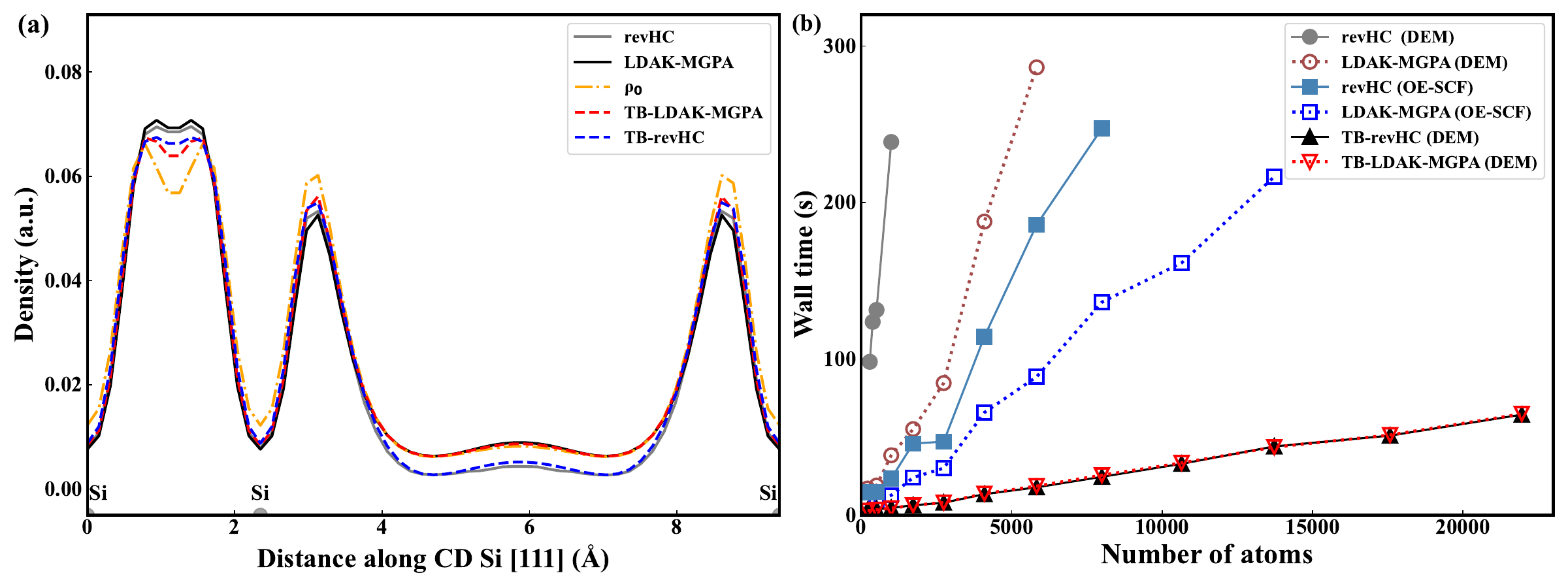}
    \caption{
    (a) Electron density distribution of CD-Si along the [111] direction, calculated using OF-DFT with revHC, LDAK-MGPA, and their TB versions, compared to the reference charge density ($\rho_0$). (b) Wall times for single-point OF-DFT calculations as a function of the number of atoms in CD-Si supercells, ranging from 216 to 21,952.
    }
    \label{fig:density}
\end{figure*}

To evaluate the reliability of the TB-KEDF scheme using the reference density ($\rho_0$), we first investigate the discrepancy of the electron densities between $\rho_0$ and the well-converged ground-state electron density $\rhor$ calculated by OF-DFT using the standard KEDFs. As shown in Fig.~\ref{fig:density}(a), the reference density ($\rho_0$) of CD-Si shows a shape roughly consistent with the well-optimized electron densities by OF-DFT using revHC and LDAK-MGPA along [111] direction. Quantitatively, the discrepancies of $\rho_0(\br)$ are 17.1\% and 18.0\% compared to the electron densities calculated by revHC and LDAK-MGPA, respectively, at the bonding region. These results suggest that our TB scheme, based on the superposition of atomic electron densities, achieves good accuracy within such a small range of electron density discrepancies. After the density optimization, we observed that the electron density deviations calculated by TB-KEDFs are only 3.3\% and 7.7\% compared to that by revHC and LDAK-MGPA functionals, respectively, in the bonding region. In addition, the electron densities calculated by TB-KEDFs are almost consistent with the results of original KEDFs in other non-bonding regions. These results further confirm the rationality of the reference density.
To evaluate the computational efficiency of this scheme, we performed single-point energy calculations for CD-Si supercells containing 216 to 21,952 atoms. All calculations utilized 92 cores on a node equipped with two AMD EPYC 9654 CPUs and 768 GB of RAM. As shown in Fig.\ref{fig:density}(b), OF-DFT calculations employing the TB-revHC/LDAK-MGPA were approximately 10 to 100 times faster than those using the original KEDFs with direct energy minimization (DEM). They remain about 5 to 10 times faster than those with original KEDFs within the OE-SCF approach. As mentioned earlier, this finding is not surprising since TB-KEDFs require only a one-shot calculation of the computationally expensive nonlocal part functional during the density optimization iterations. Table~\ref{tab:scf_calls} presents the number of nonlocal kinetic potential (or energy) calls during the density optimization for the CD-Si supercells. The results demonstrate that TB-revHC requires the fewest nonlocal kinetic potential (or energy) calls. Furthermore, OF-DFT calculations using TB-revHC also show the minimum number of calls for the (semi)local kinetic potentials (derived from the TF and vW KEDFs). The reduced number of functional evaluations is a key factor in the lower computational costs of TB-KEDFs, offering significant improvements in computational efficiency and highlighting the potential of the TB scheme for large-scale simulations. More density optimization iteration details for the LDAK-MGPA and TB-LDAK-MGPA KEDFs can be found in Table S1 of the Supplemental Material\cite{supp}.

\begin{figure*}[tb]
    \centering\includegraphics[width=1\linewidth]{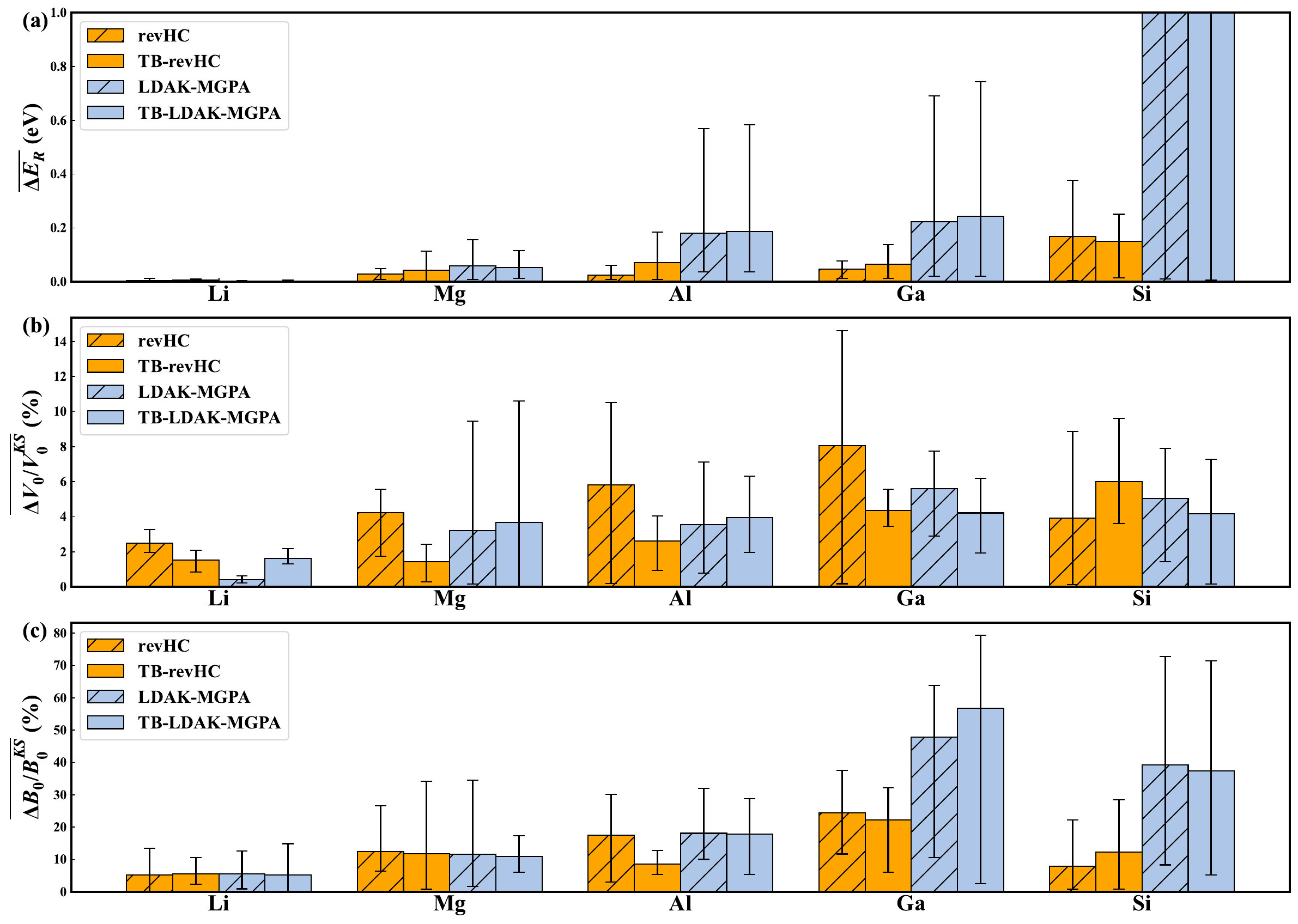}
    \caption{Comparison of the performance of KEDFs for (a) the equilibrium energy, (b) the equilibrium volume, and (c) bulk modulus. The tested KEDFs include revHC (orange with hatching), TB-revHC (orange), LDAK-MGPA (blue with hatching), and TB-LDAK-MGPA (blue). The energy error is reported as MAE in eV, while the volume and bulk modulus errors are expressed as MAPE. Error bars represent the uncertainty in the predicted values across the test.}
    \label{fig:single_ele_eos}
\end{figure*}

\begin{table}[ht]

\caption{\label{tab:scf_calls} Total number of calls for nonlocal and semilocal kinetic potential (energy) calculations for CD-Si supercells in single-point OF-DFT calculations using revHC within DEM and OE-SCF solvers, compared to those using TB-revHC within the DEM solver. The convergence criterion for the total energy is $10^{-5}$ eV/atom.}
\begin{ruledtabular}
\begin{tabular}{ccccccc}
             & \multicolumn{3}{c}{Nonlocal part calls} & \multicolumn{3}{c}{Semilocal part calls} \\ \cline{2-7} 
Number of atoms & DEM      & OE-SCF      & DEM (TB)     & DEM    & OE-SCF   & DEM (TB)   \\ \hline
8            & 314      & 30          & 1             & 314    & 600     & 68          \\
32           & 343      & 34          & 1             & 343    & 644      & 62          \\
128          & 428      & 23          & 1             & 428    & 570      & 66          \\
512          & 407      & 36          & 1             & 407    & 689      & 66          \\
1000         & 437      & 24          & 1             & 437    & 559      & 62         
\end{tabular}
\end{ruledtabular}
\end{table}

\begin{figure}
    \centering \includegraphics[width=0.6\linewidth]{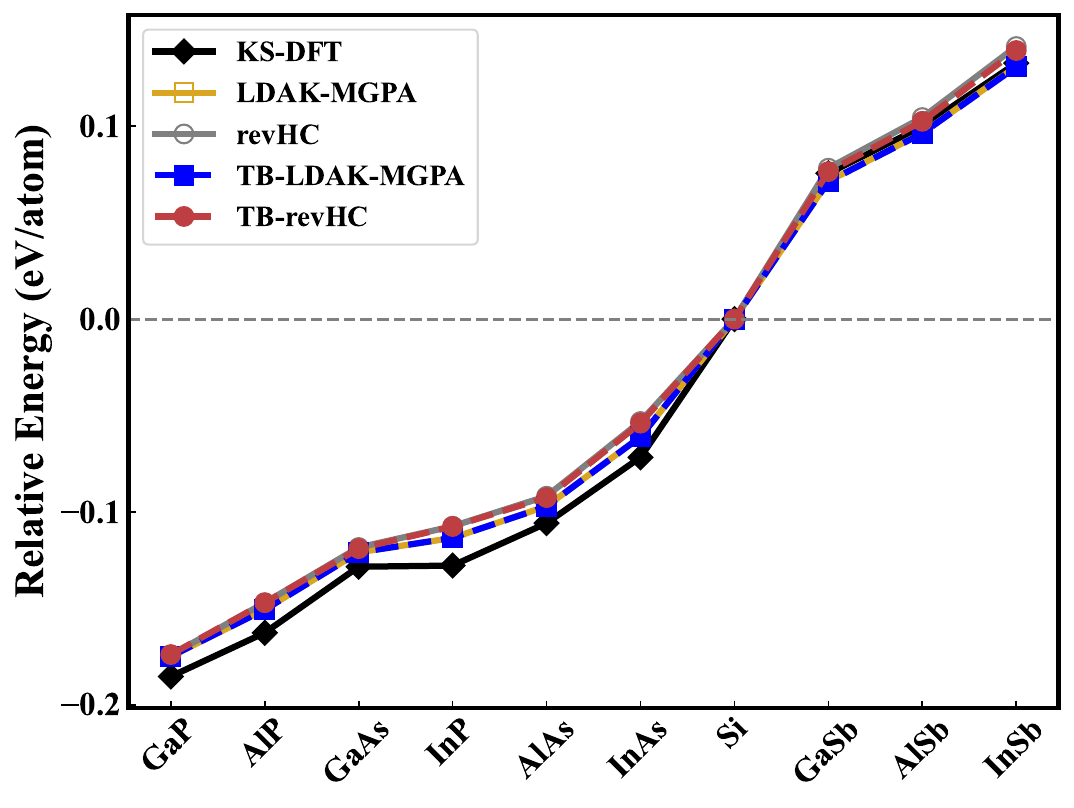}
    \caption{Relative energy differences calculated by KS-DFT and OF-DFT for nine ZB semiconductors with respect to CD-Si.}
    \label{fig:semiconductor}
\end{figure}

To verify the performance of TB-KEDFs in practical simulations, we applied them to calculate the relative equilibrium energy ($E_R$), equilibrium volume ($V_0$), and bulk modulus ($B_0$) for solid systems using Murnaghan's equation of state \cite{Murnaghan1944Compressibility}. In Fig.\ref{fig:single_ele_eos}, we present the mean absolute errors (MAEs) for $E_R$ and the mean absolute percentage errors (MAPEs) for $V_0$ and $B_0$ compared to KS-DFT results for various crystal structures. The bulk properties calculated by TB-KEDFs achieve comparable numerical accuracy to the original KEDFs, with minimal dependence on the specific element. TB-KEDFs offer comparable accuracy with the original KEDFs, providing a highly cost-effective balance between efficiency and accuracy. The data set in Fig.\ref{fig:single_ele_eos} can be found in Tables S2–S7 of the Supplemental Material\cite{supp}.

In Fig.~\ref{fig:semiconductor}, the ground-state energies of ZB semiconductors with respect to CD-Si are calculated by OF-DFT within the (TB-)KEDFs and KS-DFT. TB-KEDFs show results nearly identical to the original KEDFs and successfully reproduce the energy ordering trends among ten semiconductors as predicted by KS-DFT.

\begin{figure*}[ht]
    \centering  \includegraphics[width=\linewidth]{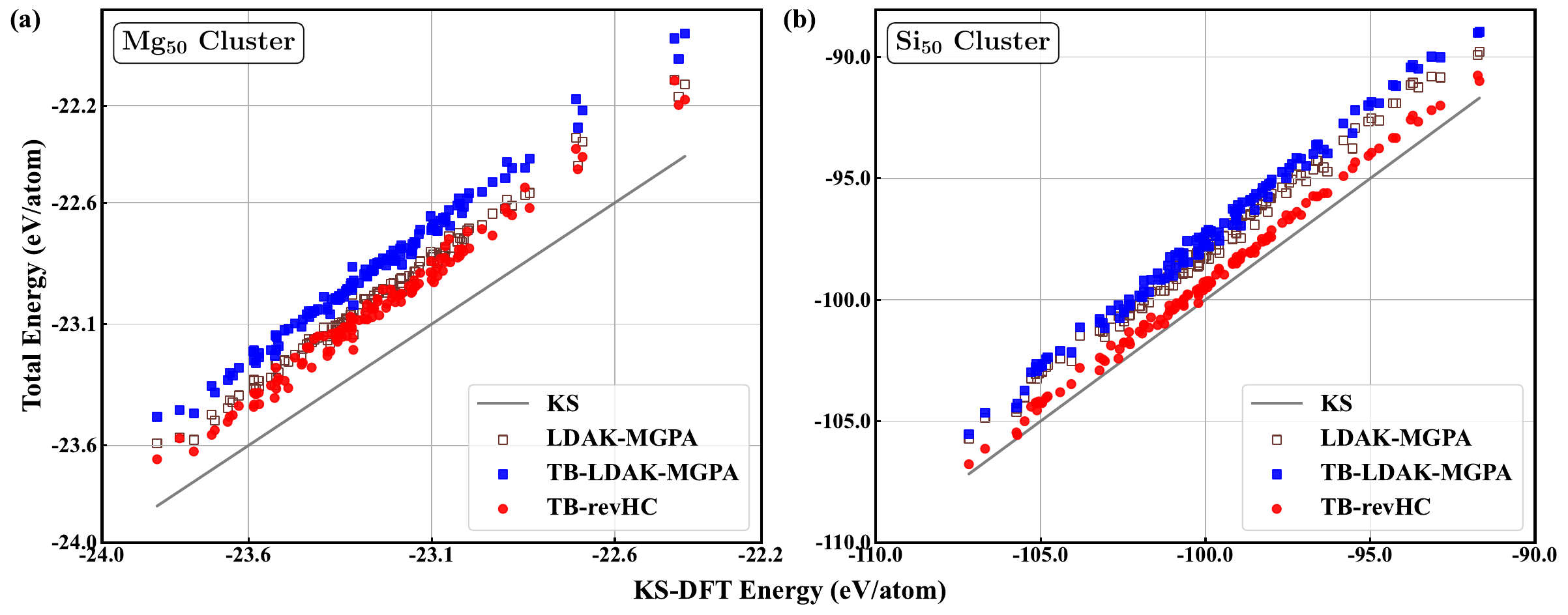}
    \caption{The energy obtained by OF-DFT with LDAK-MGPA, TB-LDAK-MGPA, TB-revHC are compared with KS-DFT for (a) Mg$_{50}$ (b) Si$_{50}$ for 120 random structures generated by CALYPSO.}
    \label{fig:cluster}
\end{figure*}

The performance of the TB scheme was also examined for clustered systems. Fig.\ref{fig:cluster} shows the total energies of the 120-random structures calculated by OF-DFT using different (TB-)KEDFs and KS-DFT. We observed that all the KEDFs generally produce trends of total energies similar to those of KS-DFT for the considered systems. OF-DFT calculations using TB-revHC especially show the closest computational results to KS-DFT. Furthermore, compared to KS-DFT results, the Pearson correlation coefficients of total energies calculated by OF-DFT using TB-revHC, TB-LDAK-MGPA, and LDAK-MGPA are 0.993, 0.996, and 0.996, respectively, for the Mg$_{50}$ random clusters; and 0.998, 0.997, and 0.997 for the Si$_{50}$ random clusters. Notably, the revHC KEDF often struggles with numerical instability, failing to converge within 100 steps for random clustered structures. The convergence details comparing revHC and TB-revHC KEDFs for the Si$_{50}$ structure are presented in Fig.~S1 of the Supplemental Material \cite{supp}. In contrast, TB-revHC achieves high numerical accuracy and demonstrates excellent numerical stability, with total energies converging to within $10^{-5}$~eV/atom after an average of 12.6 steps for all random clustered structures. These results highlight that TB-KEDFs maintain the accuracy of the original KEDFs while offering superior numerical stability and efficiency in practical simulations.

\section{Conclusions and Perspective} \label{Conclusions}
In conclusion, we have proposed an efficient framework for reconstructing nonlocal KEDFs within OF-DFT, enabling advanced KEDFs to be adopted for large-scale electronic structure calculations at a significantly reduced computational cost. The core of our approach, inspired by the tight-binding (TB) method, reconstructs the TB-KEDFs through a first-order functional expansion based on a superposition of free-atom electron densities. This strategy allows the computationally expensive nonlocal kinetic potential to be calculated only once during the electron density optimization process. Our framework offers two major advantages: (1) dramatically reducing computational cost by several orders of magnitude and (2) enhanced numerical stability during self-consistent optimization.

We demonstrated this framework's accuracy and computational efficiency by applying it to various systems, including simple metals, group III-V semiconductors, and finite systems, using advanced nonlocal KEDFs (e.g., LDAK-MGPA and revHC). Remarkably, our results show that the reconstructed TB-KEDFs nearly exactly reproduce the accuracy of the original KEDFs while achieving several orders of magnitude reduction in computational cost across all systems in OF-DFT calculations. Additionally, the reconstructed TB-KEDFs exhibit substantially improved numerical stability during electron density optimization compared to the original KEDFs.

Despite its successes, the framework's effectiveness depends on maintaining a small difference between the reference and optimized electron densities. Therefore, further refinements are both necessary and feasible. The preliminary results highlight three promising directions for extending this framework, which are the focus of our ongoing research: (1) developing more sophisticated nonlocal KEDFs with kernels beyond (semi)local density dependence to enhance accuracy and generality; (2) improving the reference electron density by using nonlocal pseudopotentials to expand the range of elements accessible in OF-DFT simulations; and (3) enhancing the TB framework with higher-order many-body corrections, akin to conventional TB methods, to improve computational accuracy. These developments hold significant potential to advance OF-DFT's accuracy, scalability, and applicability, enabling more realistic and computationally efficient materials simulations.
 



\section*{Acknowledgments}
This research was supported by the National Natural Science Foundation of China under Grants Nos. (12305002, 12274171, 12274174, T2225013, and 12174142,12034009), the National Key R\&D Program of China (Grant No. 2023YFA1406200), the Program for JLU Science and Technology Innovative Research Team. Part of the calculation was performed in the high-performance computing center of Jilin University.

\bibliography{Refs}

\begin{thebibliography}{52}%
\makeatletter
\providecommand \@ifxundefined [1]{%
 \@ifx{#1\undefined}
}%
\providecommand \@ifnum [1]{%
 \ifnum #1\expandafter \@firstoftwo
 \else \expandafter \@secondoftwo
 \fi
}%
\providecommand \@ifx [1]{%
 \ifx #1\expandafter \@firstoftwo
 \else \expandafter \@secondoftwo
 \fi
}%
\providecommand \natexlab [1]{#1}%
\providecommand \enquote  [1]{``#1''}%
\providecommand \bibnamefont  [1]{#1}%
\providecommand \bibfnamefont [1]{#1}%
\providecommand \citenamefont [1]{#1}%
\providecommand \href@noop [0]{\@secondoftwo}%
\providecommand \href [0]{\begingroup \@sanitize@url \@href}%
\providecommand \@href[1]{\@@startlink{#1}\@@href}%
\providecommand \@@href[1]{\endgroup#1\@@endlink}%
\providecommand \@sanitize@url [0]{\catcode `\\12\catcode `\$12\catcode
  `\&12\catcode `\#12\catcode `\^12\catcode `\_12\catcode `\%12\relax}%
\providecommand \@@startlink[1]{}%
\providecommand \@@endlink[0]{}%
\providecommand \url  [0]{\begingroup\@sanitize@url \@url }%
\providecommand \@url [1]{\endgroup\@href {#1}{\urlprefix }}%
\providecommand \urlprefix  [0]{URL }%
\providecommand \Eprint [0]{\href }%
\providecommand \doibase [0]{https://doi.org/}%
\providecommand \selectlanguage [0]{\@gobble}%
\providecommand \bibinfo  [0]{\@secondoftwo}%
\providecommand \bibfield  [0]{\@secondoftwo}%
\providecommand \translation [1]{[#1]}%
\providecommand \BibitemOpen [0]{}%
\providecommand \bibitemStop [0]{}%
\providecommand \bibitemNoStop [0]{.\EOS\space}%
\providecommand \EOS [0]{\spacefactor3000\relax}%
\providecommand \BibitemShut  [1]{\csname bibitem#1\endcsname}%
\let\auto@bib@innerbib\@empty
\bibitem [{\citenamefont {Hohenberg}\ and\ \citenamefont
  {Kohn}(1964)}]{hk1964}%
  \BibitemOpen
  \bibfield  {author} {\bibinfo {author} {\bibfnamefont {P.}~\bibnamefont
  {Hohenberg}}\ and\ \bibinfo {author} {\bibfnamefont {W.}~\bibnamefont
  {Kohn}},\ }\bibfield  {title} {\bibinfo {title} {Inhomogeneous electron
  gas},\ }\href {https://doi.org/10.1103/PhysRev.136.B864} {\bibfield
  {journal} {\bibinfo  {journal} {Phys. Rev.}\ }\textbf {\bibinfo {volume}
  {136}},\ \bibinfo {pages} {B864} (\bibinfo {year} {1964})}\BibitemShut
  {NoStop}%
\bibitem [{\citenamefont {Kohn}\ and\ \citenamefont
  {Sham}(1965)}]{kohn1965self}%
  \BibitemOpen
  \bibfield  {author} {\bibinfo {author} {\bibfnamefont {W.}~\bibnamefont
  {Kohn}}\ and\ \bibinfo {author} {\bibfnamefont {L.~J.}\ \bibnamefont
  {Sham}},\ }\bibfield  {title} {\bibinfo {title} {Self-consistent equations
  including exchange and correlation effects},\ }\href
  {https://doi.org/10.1103/PhysRev.140.A1133} {\bibfield  {journal} {\bibinfo
  {journal} {Phys. Rev.}\ }\textbf {\bibinfo {volume} {140}},\ \bibinfo {pages}
  {A1133} (\bibinfo {year} {1965})}\BibitemShut {NoStop}%
\bibitem [{\citenamefont {Carloni}\ \emph {et~al.}(2002)\citenamefont
  {Carloni}, \citenamefont {Rothlisberger},\ and\ \citenamefont
  {Parrinello}}]{carloni2002role}%
  \BibitemOpen
  \bibfield  {author} {\bibinfo {author} {\bibfnamefont {P.}~\bibnamefont
  {Carloni}}, \bibinfo {author} {\bibfnamefont {U.}~\bibnamefont
  {Rothlisberger}},\ and\ \bibinfo {author} {\bibfnamefont {M.}~\bibnamefont
  {Parrinello}},\ }\bibfield  {title} {\bibinfo {title} {The role and
  perspective of ab initio molecular dynamics in the study of biological
  systems},\ }\href@noop {} {\bibfield  {journal} {\bibinfo  {journal} {Acc.
  Chem. Res.}\ }\textbf {\bibinfo {volume} {35}},\ \bibinfo {pages} {455}
  (\bibinfo {year} {2002})}\BibitemShut {NoStop}%
\bibitem [{\citenamefont {Burke}(2012)}]{burke2012perspective}%
  \BibitemOpen
  \bibfield  {author} {\bibinfo {author} {\bibfnamefont {K.}~\bibnamefont
  {Burke}},\ }\bibfield  {title} {\bibinfo {title} {Perspective on density
  functional theory},\ }\href@noop {} {\bibfield  {journal} {\bibinfo
  {journal} {J. Chem. Phys.}\ }\textbf {\bibinfo {volume} {136}} (\bibinfo
  {year} {2012})}\BibitemShut {NoStop}%
\bibitem [{\citenamefont {Karasiev}\ and\ \citenamefont
  {Trickey}(2012)}]{karasiev2012issues}%
  \BibitemOpen
  \bibfield  {author} {\bibinfo {author} {\bibfnamefont {V.~V.}\ \bibnamefont
  {Karasiev}}\ and\ \bibinfo {author} {\bibfnamefont {S.~B.}\ \bibnamefont
  {Trickey}},\ }\bibfield  {title} {\bibinfo {title} {Issues and challenges in
  orbital-free density functional calculations},\ }\href
  {https://doi.org/https://doi.org/10.1016/j.cpc.2012.06.016} {\bibfield
  {journal} {\bibinfo  {journal} {Comput. Phys. Commun.}\ }\textbf {\bibinfo
  {volume} {183}},\ \bibinfo {pages} {2519} (\bibinfo {year}
  {2012})}\BibitemShut {NoStop}%
\bibitem [{\citenamefont {Witt}\ \emph {et~al.}(2018)\citenamefont {Witt},
  \citenamefont {Del~Rio}, \citenamefont {Dieterich},\ and\ \citenamefont
  {Carter}}]{witt2018orbital}%
  \BibitemOpen
  \bibfield  {author} {\bibinfo {author} {\bibfnamefont {W.~C.}\ \bibnamefont
  {Witt}}, \bibinfo {author} {\bibfnamefont {B.~G.}\ \bibnamefont {Del~Rio}},
  \bibinfo {author} {\bibfnamefont {J.~M.}\ \bibnamefont {Dieterich}},\ and\
  \bibinfo {author} {\bibfnamefont {E.~A.}\ \bibnamefont {Carter}},\ }\bibfield
   {title} {\bibinfo {title} {Orbital-free density functional theory for
  materials research},\ }\href {https://doi.org/10.1557/jmr.2017.462}
  {\bibfield  {journal} {\bibinfo  {journal} {J. Mater. Res.}\ }\textbf
  {\bibinfo {volume} {33}},\ \bibinfo {pages} {777} (\bibinfo {year}
  {2018})}\BibitemShut {NoStop}%
\bibitem [{\citenamefont {Mi}\ \emph {et~al.}(2023)\citenamefont {Mi},
  \citenamefont {Luo}, \citenamefont {Trickey},\ and\ \citenamefont
  {Pavanello}}]{mi2023}%
  \BibitemOpen
  \bibfield  {author} {\bibinfo {author} {\bibfnamefont {W.}~\bibnamefont
  {Mi}}, \bibinfo {author} {\bibfnamefont {K.}~\bibnamefont {Luo}}, \bibinfo
  {author} {\bibfnamefont {S.~B.}\ \bibnamefont {Trickey}},\ and\ \bibinfo
  {author} {\bibfnamefont {M.}~\bibnamefont {Pavanello}},\ }\bibfield  {title}
  {\bibinfo {title} {Orbital-free density functional theory: An attractive
  electronic structure method for large-scale first-principles simulations},\
  }\href {https://doi.org/10.1021/acs.chemrev.2c00758} {\bibfield  {journal}
  {\bibinfo  {journal} {Chem. Rev.}\ }\textbf {\bibinfo {volume} {123}},\
  \bibinfo {pages} {12039} (\bibinfo {year} {2023})}\BibitemShut {NoStop}%
\bibitem [{\citenamefont {Xu}\ \emph {et~al.}(2024)\citenamefont {Xu},
  \citenamefont {Ma}, \citenamefont {Mi}, \citenamefont {Wang},\ and\
  \citenamefont {Ma}}]{xu2024recent}%
  \BibitemOpen
  \bibfield  {author} {\bibinfo {author} {\bibfnamefont {Q.}~\bibnamefont
  {Xu}}, \bibinfo {author} {\bibfnamefont {C.}~\bibnamefont {Ma}}, \bibinfo
  {author} {\bibfnamefont {W.}~\bibnamefont {Mi}}, \bibinfo {author}
  {\bibfnamefont {Y.}~\bibnamefont {Wang}},\ and\ \bibinfo {author}
  {\bibfnamefont {Y.}~\bibnamefont {Ma}},\ }\bibfield  {title} {\bibinfo
  {title} {Recent advancements and challenges in orbital-free density
  functional theory},\ }\href@noop {} {\bibfield  {journal} {\bibinfo
  {journal} {WIREs Comput. Mol. Sci.}\ }\textbf {\bibinfo {volume} {14}},\
  \bibinfo {pages} {e1724} (\bibinfo {year} {2024})}\BibitemShut {NoStop}%
\bibitem [{\citenamefont {Shao}\ \emph {et~al.}(2018)\citenamefont {Shao},
  \citenamefont {Xu}, \citenamefont {Wang}, \citenamefont {Lv}, \citenamefont
  {Wang},\ and\ \citenamefont {Ma}}]{shao2018large}%
  \BibitemOpen
  \bibfield  {author} {\bibinfo {author} {\bibfnamefont {X.}~\bibnamefont
  {Shao}}, \bibinfo {author} {\bibfnamefont {Q.}~\bibnamefont {Xu}}, \bibinfo
  {author} {\bibfnamefont {S.}~\bibnamefont {Wang}}, \bibinfo {author}
  {\bibfnamefont {J.}~\bibnamefont {Lv}}, \bibinfo {author} {\bibfnamefont
  {Y.}~\bibnamefont {Wang}},\ and\ \bibinfo {author} {\bibfnamefont
  {Y.}~\bibnamefont {Ma}},\ }\bibfield  {title} {\bibinfo {title} {Large-scale
  ab initio simulations for periodic system},\ }\href@noop {} {\bibfield
  {journal} {\bibinfo  {journal} {Comput. Phys. Commun.}\ }\textbf {\bibinfo
  {volume} {233}},\ \bibinfo {pages} {78} (\bibinfo {year} {2018})}\BibitemShut
  {NoStop}%
\bibitem [{\citenamefont {Hung}\ and\ \citenamefont
  {Carter}(2009)}]{hung2009accurate}%
  \BibitemOpen
  \bibfield  {author} {\bibinfo {author} {\bibfnamefont {L.}~\bibnamefont
  {Hung}}\ and\ \bibinfo {author} {\bibfnamefont {E.~A.}\ \bibnamefont
  {Carter}},\ }\bibfield  {title} {\bibinfo {title} {Accurate simulations of
  metals at the mesoscale: Explicit treatment of 1 million atoms with quantum
  mechanics},\ }\href@noop {} {\bibfield  {journal} {\bibinfo  {journal} {Chem.
  Phys. Lett.}\ }\textbf {\bibinfo {volume} {475}},\ \bibinfo {pages} {163}
  (\bibinfo {year} {2009})}\BibitemShut {NoStop}%
\bibitem [{\citenamefont {Shao}\ \emph
  {et~al.}(2021{\natexlab{a}})\citenamefont {Shao}, \citenamefont {Jiang},
  \citenamefont {Mi}, \citenamefont {Genova},\ and\ \citenamefont
  {Pavanello}}]{shao2021dftpy}%
  \BibitemOpen
  \bibfield  {author} {\bibinfo {author} {\bibfnamefont {X.}~\bibnamefont
  {Shao}}, \bibinfo {author} {\bibfnamefont {K.}~\bibnamefont {Jiang}},
  \bibinfo {author} {\bibfnamefont {W.}~\bibnamefont {Mi}}, \bibinfo {author}
  {\bibfnamefont {A.}~\bibnamefont {Genova}},\ and\ \bibinfo {author}
  {\bibfnamefont {M.}~\bibnamefont {Pavanello}},\ }\bibfield  {title} {\bibinfo
  {title} {Dftpy: An efficient and object-oriented platform for orbital-free
  dft simulations},\ }\href@noop {} {\bibfield  {journal} {\bibinfo  {journal}
  {WIREs Comput. Mol. Sci.}\ }\textbf {\bibinfo {volume} {11}},\ \bibinfo
  {pages} {e1482} (\bibinfo {year} {2021}{\natexlab{a}})}\BibitemShut {NoStop}%
\bibitem [{\citenamefont {Weizs{\"a}cker}(1935)}]{weizsacker1935}%
  \BibitemOpen
  \bibfield  {author} {\bibinfo {author} {\bibfnamefont {C.~F.~v.}\
  \bibnamefont {Weizs{\"a}cker}},\ }\bibfield  {title} {\bibinfo {title} {Zur
  theorie der kernmassen},\ }\href {https://doi.org/10.1007/BF01337700}
  {\bibfield  {journal} {\bibinfo  {journal} {Z. Phys.}\ }\textbf {\bibinfo
  {volume} {96}},\ \bibinfo {pages} {431} (\bibinfo {year} {1935})}\BibitemShut
  {NoStop}%
\bibitem [{\citenamefont {Laricchia}\ \emph {et~al.}(2011)\citenamefont
  {Laricchia}, \citenamefont {Fabiano}, \citenamefont {Constantin},\ and\
  \citenamefont {Della~Sala}}]{laricchia2011generalized}%
  \BibitemOpen
  \bibfield  {author} {\bibinfo {author} {\bibfnamefont {S.}~\bibnamefont
  {Laricchia}}, \bibinfo {author} {\bibfnamefont {E.}~\bibnamefont {Fabiano}},
  \bibinfo {author} {\bibfnamefont {L.}~\bibnamefont {Constantin}},\ and\
  \bibinfo {author} {\bibfnamefont {F.}~\bibnamefont {Della~Sala}},\ }\bibfield
   {title} {\bibinfo {title} {Generalized gradient approximations of the
  noninteracting kinetic energy from the semiclassical atom theory:
  Rationalization of the accuracy of the frozen density embedding theory for
  nonbonded interactions},\ }\href@noop {} {\bibfield  {journal} {\bibinfo
  {journal} {J. Chem. Theory Comput.}\ }\textbf {\bibinfo {volume} {7}},\
  \bibinfo {pages} {2439} (\bibinfo {year} {2011})}\BibitemShut {NoStop}%
\bibitem [{\citenamefont {Ou-Yang}\ and\ \citenamefont
  {Levy}(1991)}]{ou1991approximate}%
  \BibitemOpen
  \bibfield  {author} {\bibinfo {author} {\bibfnamefont {H.}~\bibnamefont
  {Ou-Yang}}\ and\ \bibinfo {author} {\bibfnamefont {M.}~\bibnamefont {Levy}},\
  }\bibfield  {title} {\bibinfo {title} {Approximate noninteracting kinetic
  energy functionals from a nonuniform scaling requirement},\ }\href@noop {}
  {\bibfield  {journal} {\bibinfo  {journal} {Int. J. Quantum Chem.}\ }\textbf
  {\bibinfo {volume} {40}},\ \bibinfo {pages} {379} (\bibinfo {year}
  {1991})}\BibitemShut {NoStop}%
\bibitem [{\citenamefont {Perdew}(1992)}]{perdew1992generalized}%
  \BibitemOpen
  \bibfield  {author} {\bibinfo {author} {\bibfnamefont {J.~P.}\ \bibnamefont
  {Perdew}},\ }\bibfield  {title} {\bibinfo {title} {Generalized gradient
  approximation for the fermion kinetic energy as a functional of the
  density},\ }\href@noop {} {\bibfield  {journal} {\bibinfo  {journal} {Phys.
  Lett. A}\ }\textbf {\bibinfo {volume} {165}},\ \bibinfo {pages} {79}
  (\bibinfo {year} {1992})}\BibitemShut {NoStop}%
\bibitem [{\citenamefont {Ernzerhof}(2000)}]{ernzerhof2000role}%
  \BibitemOpen
  \bibfield  {author} {\bibinfo {author} {\bibfnamefont {M.}~\bibnamefont
  {Ernzerhof}},\ }\bibfield  {title} {\bibinfo {title} {The role of the kinetic
  energy density in approximations to the exchange energy},\ }\href@noop {}
  {\bibfield  {journal} {\bibinfo  {journal} {J. Mol. Struct.: THEOCHEM}\
  }\textbf {\bibinfo {volume} {501}},\ \bibinfo {pages} {59} (\bibinfo {year}
  {2000})}\BibitemShut {NoStop}%
\bibitem [{\citenamefont {Constantin}\ and\ \citenamefont
  {Ruzsinszky}(2009)}]{constantin2009kinetic}%
  \BibitemOpen
  \bibfield  {author} {\bibinfo {author} {\bibfnamefont {L.~A.}\ \bibnamefont
  {Constantin}}\ and\ \bibinfo {author} {\bibfnamefont {A.}~\bibnamefont
  {Ruzsinszky}},\ }\bibfield  {title} {\bibinfo {title} {Kinetic energy density
  functionals from the airy gas with an application to the atomization kinetic
  energies of molecules},\ }\href@noop {} {\bibfield  {journal} {\bibinfo
  {journal} {Phys. Rev. B}\ }\textbf {\bibinfo {volume} {79}},\ \bibinfo
  {pages} {115117} (\bibinfo {year} {2009})}\BibitemShut {NoStop}%
\bibitem [{\citenamefont {Vitos}\ \emph {et~al.}(2000)\citenamefont {Vitos},
  \citenamefont {Johansson}, \citenamefont {Kollar},\ and\ \citenamefont
  {Skriver}}]{vitos2000local}%
  \BibitemOpen
  \bibfield  {author} {\bibinfo {author} {\bibfnamefont {L.}~\bibnamefont
  {Vitos}}, \bibinfo {author} {\bibfnamefont {B.}~\bibnamefont {Johansson}},
  \bibinfo {author} {\bibfnamefont {J.}~\bibnamefont {Kollar}},\ and\ \bibinfo
  {author} {\bibfnamefont {H.~L.}\ \bibnamefont {Skriver}},\ }\bibfield
  {title} {\bibinfo {title} {Local kinetic-energy density of the airy gas},\
  }\href@noop {} {\bibfield  {journal} {\bibinfo  {journal} {Phys. Rev. A}\
  }\textbf {\bibinfo {volume} {61}},\ \bibinfo {pages} {052511} (\bibinfo
  {year} {2000})}\BibitemShut {NoStop}%
\bibitem [{\citenamefont {Thakkar}(1992)}]{thakkar1992comparison}%
  \BibitemOpen
  \bibfield  {author} {\bibinfo {author} {\bibfnamefont {A.~J.}\ \bibnamefont
  {Thakkar}},\ }\bibfield  {title} {\bibinfo {title} {Comparison of
  kinetic-energy density functionals},\ }\href@noop {} {\bibfield  {journal}
  {\bibinfo  {journal} {Phys. Rev. A}\ }\textbf {\bibinfo {volume} {46}},\
  \bibinfo {pages} {6920} (\bibinfo {year} {1992})}\BibitemShut {NoStop}%
\bibitem [{\citenamefont {Lembarki}\ and\ \citenamefont
  {Chermette}(1994)}]{lembarki1994obtaining}%
  \BibitemOpen
  \bibfield  {author} {\bibinfo {author} {\bibfnamefont {A.}~\bibnamefont
  {Lembarki}}\ and\ \bibinfo {author} {\bibfnamefont {H.}~\bibnamefont
  {Chermette}},\ }\bibfield  {title} {\bibinfo {title} {Obtaining a
  gradient-corrected kinetic-energy functional from the perdew-wang exchange
  functional},\ }\href@noop {} {\bibfield  {journal} {\bibinfo  {journal}
  {Phys. Rev. A}\ }\textbf {\bibinfo {volume} {50}},\ \bibinfo {pages} {5328}
  (\bibinfo {year} {1994})}\BibitemShut {NoStop}%
\bibitem [{\citenamefont {Luo}\ \emph {et~al.}(2018)\citenamefont {Luo},
  \citenamefont {Karasiev},\ and\ \citenamefont {Trickey}}]{luo2018simple}%
  \BibitemOpen
  \bibfield  {author} {\bibinfo {author} {\bibfnamefont {K.}~\bibnamefont
  {Luo}}, \bibinfo {author} {\bibfnamefont {V.~V.}\ \bibnamefont {Karasiev}},\
  and\ \bibinfo {author} {\bibfnamefont {S.}~\bibnamefont {Trickey}},\
  }\bibfield  {title} {\bibinfo {title} {A simple generalized gradient
  approximation for the noninteracting kinetic energy density functional},\
  }\href@noop {} {\bibfield  {journal} {\bibinfo  {journal} {Phys. Rev. B}\
  }\textbf {\bibinfo {volume} {98}},\ \bibinfo {pages} {041111} (\bibinfo
  {year} {2018})}\BibitemShut {NoStop}%
\bibitem [{\citenamefont {Garcia-Aldea}\ and\ \citenamefont
  {Alvarellos}(2007)}]{garcia2007kinetic}%
  \BibitemOpen
  \bibfield  {author} {\bibinfo {author} {\bibfnamefont {D.}~\bibnamefont
  {Garcia-Aldea}}\ and\ \bibinfo {author} {\bibfnamefont {J.}~\bibnamefont
  {Alvarellos}},\ }\bibfield  {title} {\bibinfo {title} {Kinetic-energy density
  functionals with nonlocal terms with the structure of the thomas-fermi
  functional},\ }\href@noop {} {\bibfield  {journal} {\bibinfo  {journal}
  {Phys. Rev. A}\ }\textbf {\bibinfo {volume} {76}},\ \bibinfo {pages} {052504}
  (\bibinfo {year} {2007})}\BibitemShut {NoStop}%
\bibitem [{\citenamefont {Chac{\'o}n}\ \emph {et~al.}(1985)\citenamefont
  {Chac{\'o}n}, \citenamefont {Alvarellos},\ and\ \citenamefont
  {Tarazona}}]{chacon1985nonlocal}%
  \BibitemOpen
  \bibfield  {author} {\bibinfo {author} {\bibfnamefont {E.}~\bibnamefont
  {Chac{\'o}n}}, \bibinfo {author} {\bibfnamefont {J.}~\bibnamefont
  {Alvarellos}},\ and\ \bibinfo {author} {\bibfnamefont {P.}~\bibnamefont
  {Tarazona}},\ }\bibfield  {title} {\bibinfo {title} {Nonlocal kinetic energy
  functional for nonhomogeneous electron systems},\ }\href@noop {} {\bibfield
  {journal} {\bibinfo  {journal} {Phys. Rev. B}\ }\textbf {\bibinfo {volume}
  {32}},\ \bibinfo {pages} {7868} (\bibinfo {year} {1985})}\BibitemShut
  {NoStop}%
\bibitem [{\citenamefont {Wang}\ and\ \citenamefont
  {Teter}(1992)}]{wang1992kinetic}%
  \BibitemOpen
  \bibfield  {author} {\bibinfo {author} {\bibfnamefont {L.-W.}\ \bibnamefont
  {Wang}}\ and\ \bibinfo {author} {\bibfnamefont {M.~P.}\ \bibnamefont
  {Teter}},\ }\bibfield  {title} {\bibinfo {title} {Kinetic-energy functional
  of the electron density},\ }\href@noop {} {\bibfield  {journal} {\bibinfo
  {journal} {Phys. Rev. B}\ }\textbf {\bibinfo {volume} {45}},\ \bibinfo
  {pages} {13196} (\bibinfo {year} {1992})}\BibitemShut {NoStop}%
\bibitem [{\citenamefont {Smargiassi}\ and\ \citenamefont
  {Madden}(1994)}]{smargiassi1994orbital}%
  \BibitemOpen
  \bibfield  {author} {\bibinfo {author} {\bibfnamefont {E.}~\bibnamefont
  {Smargiassi}}\ and\ \bibinfo {author} {\bibfnamefont {P.~A.}\ \bibnamefont
  {Madden}},\ }\bibfield  {title} {\bibinfo {title} {Orbital-free
  kinetic-energy functionals for first-principles molecular dynamics},\
  }\href@noop {} {\bibfield  {journal} {\bibinfo  {journal} {Phys. Rev. B}\
  }\textbf {\bibinfo {volume} {49}},\ \bibinfo {pages} {5220} (\bibinfo {year}
  {1994})}\BibitemShut {NoStop}%
\bibitem [{\citenamefont {Perrot}(1994)}]{perrot1994hydrogen}%
  \BibitemOpen
  \bibfield  {author} {\bibinfo {author} {\bibfnamefont {F.}~\bibnamefont
  {Perrot}},\ }\bibfield  {title} {\bibinfo {title} {Hydrogen-hydrogen
  interaction in an electron gas},\ }\href@noop {} {\bibfield  {journal}
  {\bibinfo  {journal} {J. Phys.: Condens. Matter}\ }\textbf {\bibinfo {volume}
  {6}},\ \bibinfo {pages} {431} (\bibinfo {year} {1994})}\BibitemShut {NoStop}%
\bibitem [{\citenamefont {Wang}\ \emph {et~al.}(1999)\citenamefont {Wang},
  \citenamefont {Govind},\ and\ \citenamefont {Carter}}]{wang1999orbital}%
  \BibitemOpen
  \bibfield  {author} {\bibinfo {author} {\bibfnamefont {Y.~A.}\ \bibnamefont
  {Wang}}, \bibinfo {author} {\bibfnamefont {N.}~\bibnamefont {Govind}},\ and\
  \bibinfo {author} {\bibfnamefont {E.~A.}\ \bibnamefont {Carter}},\ }\bibfield
   {title} {\bibinfo {title} {Orbital-free kinetic-energy density functionals
  with a density-dependent kernel},\ }\href@noop {} {\bibfield  {journal}
  {\bibinfo  {journal} {Phys. Rev. B}\ }\textbf {\bibinfo {volume} {60}},\
  \bibinfo {pages} {16350} (\bibinfo {year} {1999})}\BibitemShut {NoStop}%
\bibitem [{\citenamefont {Huang}\ and\ \citenamefont
  {Carter}(2010)}]{huang2010nonlocal}%
  \BibitemOpen
  \bibfield  {author} {\bibinfo {author} {\bibfnamefont {C.}~\bibnamefont
  {Huang}}\ and\ \bibinfo {author} {\bibfnamefont {E.~A.}\ \bibnamefont
  {Carter}},\ }\bibfield  {title} {\bibinfo {title} {Nonlocal orbital-free
  kinetic energy density functional for semiconductors},\ }\href@noop {}
  {\bibfield  {journal} {\bibinfo  {journal} {Phys. Rev. B}\ }\textbf {\bibinfo
  {volume} {81}},\ \bibinfo {pages} {045206} (\bibinfo {year}
  {2010})}\BibitemShut {NoStop}%
\bibitem [{\citenamefont {Mi}\ \emph {et~al.}(2018)\citenamefont {Mi},
  \citenamefont {Genova},\ and\ \citenamefont {Pavanello}}]{mi2018nonlocal}%
  \BibitemOpen
  \bibfield  {author} {\bibinfo {author} {\bibfnamefont {W.}~\bibnamefont
  {Mi}}, \bibinfo {author} {\bibfnamefont {A.}~\bibnamefont {Genova}},\ and\
  \bibinfo {author} {\bibfnamefont {M.}~\bibnamefont {Pavanello}},\ }\bibfield
  {title} {\bibinfo {title} {Nonlocal kinetic energy functionals by functional
  integration},\ }\href@noop {} {\bibfield  {journal} {\bibinfo  {journal} {J.
  Chem. Phys.}\ }\textbf {\bibinfo {volume} {148}},\ \bibinfo {pages} {184107}
  (\bibinfo {year} {2018})}\BibitemShut {NoStop}%
\bibitem [{\citenamefont {Xu}\ \emph {et~al.}(2019)\citenamefont {Xu},
  \citenamefont {Wang},\ and\ \citenamefont {Ma}}]{xu2019nonlocal}%
  \BibitemOpen
  \bibfield  {author} {\bibinfo {author} {\bibfnamefont {Q.}~\bibnamefont
  {Xu}}, \bibinfo {author} {\bibfnamefont {Y.}~\bibnamefont {Wang}},\ and\
  \bibinfo {author} {\bibfnamefont {Y.}~\bibnamefont {Ma}},\ }\bibfield
  {title} {\bibinfo {title} {Nonlocal kinetic energy density functional via
  line integrals and its application to orbital-free density functional
  theory},\ }\href@noop {} {\bibfield  {journal} {\bibinfo  {journal} {Phys.
  Rev. B}\ }\textbf {\bibinfo {volume} {100}},\ \bibinfo {pages} {205132}
  (\bibinfo {year} {2019})}\BibitemShut {NoStop}%
\bibitem [{\citenamefont {Yang}(1987)}]{yang1987ab}%
  \BibitemOpen
  \bibfield  {author} {\bibinfo {author} {\bibfnamefont {W.}~\bibnamefont
  {Yang}},\ }\bibfield  {title} {\bibinfo {title} {Ab initio approach for
  many-electron systems without invoking orbitals: An integral formulation of
  density-functional theory},\ }\href@noop {} {\bibfield  {journal} {\bibinfo
  {journal} {Phys. Rev. Lett.}\ }\textbf {\bibinfo {volume} {59}},\ \bibinfo
  {pages} {1569} (\bibinfo {year} {1987})}\BibitemShut {NoStop}%
\bibitem [{\citenamefont {Shao}\ \emph
  {et~al.}(2021{\natexlab{b}})\citenamefont {Shao}, \citenamefont {Mi},\ and\
  \citenamefont {Pavanello}}]{shao2021revised}%
  \BibitemOpen
  \bibfield  {author} {\bibinfo {author} {\bibfnamefont {X.}~\bibnamefont
  {Shao}}, \bibinfo {author} {\bibfnamefont {W.}~\bibnamefont {Mi}},\ and\
  \bibinfo {author} {\bibfnamefont {M.}~\bibnamefont {Pavanello}},\ }\bibfield
  {title} {\bibinfo {title} {Revised huang-carter nonlocal kinetic energy
  functional for semiconductors and their surfaces},\ }\href@noop {} {\bibfield
   {journal} {\bibinfo  {journal} {Phys. Rev. B}\ }\textbf {\bibinfo {volume}
  {104}},\ \bibinfo {pages} {045118} (\bibinfo {year}
  {2021}{\natexlab{b}})}\BibitemShut {NoStop}%
\bibitem [{\citenamefont {Mi}\ and\ \citenamefont
  {Pavanello}(2019)}]{mi2019orbital}%
  \BibitemOpen
  \bibfield  {author} {\bibinfo {author} {\bibfnamefont {W.}~\bibnamefont
  {Mi}}\ and\ \bibinfo {author} {\bibfnamefont {M.}~\bibnamefont {Pavanello}},\
  }\bibfield  {title} {\bibinfo {title} {Orbital-free density functional theory
  correctly models quantum dots when asymptotics, nonlocality, and
  nonhomogeneity are accounted for},\ }\href@noop {} {\bibfield  {journal}
  {\bibinfo  {journal} {Phys. Rev. B}\ }\textbf {\bibinfo {volume} {100}},\
  \bibinfo {pages} {041105} (\bibinfo {year} {2019})}\BibitemShut {NoStop}%
\bibitem [{\citenamefont {Xu}\ \emph {et~al.}(2020)\citenamefont {Xu},
  \citenamefont {Lv}, \citenamefont {Wang},\ and\ \citenamefont
  {Ma}}]{xu2020nonlocal}%
  \BibitemOpen
  \bibfield  {author} {\bibinfo {author} {\bibfnamefont {Q.}~\bibnamefont
  {Xu}}, \bibinfo {author} {\bibfnamefont {J.}~\bibnamefont {Lv}}, \bibinfo
  {author} {\bibfnamefont {Y.}~\bibnamefont {Wang}},\ and\ \bibinfo {author}
  {\bibfnamefont {Y.}~\bibnamefont {Ma}},\ }\bibfield  {title} {\bibinfo
  {title} {Nonlocal kinetic energy density functionals for isolated systems
  obtained via local density approximation kernels},\ }\href@noop {} {\bibfield
   {journal} {\bibinfo  {journal} {Phys. Rev. B}\ }\textbf {\bibinfo {volume}
  {101}},\ \bibinfo {pages} {045110} (\bibinfo {year} {2020})}\BibitemShut
  {NoStop}%
\bibitem [{\citenamefont {Constantin}\ \emph {et~al.}(2018)\citenamefont
  {Constantin}, \citenamefont {Fabiano},\ and\ \citenamefont
  {Della~Sala}}]{constantin2018nonlocal}%
  \BibitemOpen
  \bibfield  {author} {\bibinfo {author} {\bibfnamefont {L.~A.}\ \bibnamefont
  {Constantin}}, \bibinfo {author} {\bibfnamefont {E.}~\bibnamefont
  {Fabiano}},\ and\ \bibinfo {author} {\bibfnamefont {F.}~\bibnamefont
  {Della~Sala}},\ }\bibfield  {title} {\bibinfo {title} {Nonlocal kinetic
  energy functional from the jellium-with-gap model: Applications to
  orbital-free density functional theory},\ }\href@noop {} {\bibfield
  {journal} {\bibinfo  {journal} {Phys. Rev. B}\ }\textbf {\bibinfo {volume}
  {97}},\ \bibinfo {pages} {205137} (\bibinfo {year} {2018})}\BibitemShut
  {NoStop}%
\bibitem [{\citenamefont {Shao}\ \emph
  {et~al.}(2021{\natexlab{c}})\citenamefont {Shao}, \citenamefont {Mi},\ and\
  \citenamefont {Pavanello}}]{shao2021efficient}%
  \BibitemOpen
  \bibfield  {author} {\bibinfo {author} {\bibfnamefont {X.}~\bibnamefont
  {Shao}}, \bibinfo {author} {\bibfnamefont {W.}~\bibnamefont {Mi}},\ and\
  \bibinfo {author} {\bibfnamefont {M.}~\bibnamefont {Pavanello}},\ }\bibfield
  {title} {\bibinfo {title} {Efficient dft solver for nanoscale simulations and
  beyond},\ }\href@noop {} {\bibfield  {journal} {\bibinfo  {journal} {J. Phys.
  Chem. Lett.}\ }\textbf {\bibinfo {volume} {12}},\ \bibinfo {pages} {4134}
  (\bibinfo {year} {2021}{\natexlab{c}})}\BibitemShut {NoStop}%
\bibitem [{\citenamefont {Elstner}\ \emph {et~al.}(1998)\citenamefont
  {Elstner}, \citenamefont {Porezag}, \citenamefont {Jungnickel}, \citenamefont
  {Elsner}, \citenamefont {Haugk}, \citenamefont {Frauenheim}, \citenamefont
  {Suhai},\ and\ \citenamefont {Seifert}}]{elstner1998self}%
  \BibitemOpen
  \bibfield  {author} {\bibinfo {author} {\bibfnamefont {M.}~\bibnamefont
  {Elstner}}, \bibinfo {author} {\bibfnamefont {D.}~\bibnamefont {Porezag}},
  \bibinfo {author} {\bibfnamefont {G.}~\bibnamefont {Jungnickel}}, \bibinfo
  {author} {\bibfnamefont {J.}~\bibnamefont {Elsner}}, \bibinfo {author}
  {\bibfnamefont {M.}~\bibnamefont {Haugk}}, \bibinfo {author} {\bibfnamefont
  {T.}~\bibnamefont {Frauenheim}}, \bibinfo {author} {\bibfnamefont
  {S.}~\bibnamefont {Suhai}},\ and\ \bibinfo {author} {\bibfnamefont
  {G.}~\bibnamefont {Seifert}},\ }\bibfield  {title} {\bibinfo {title}
  {Self-consistent-charge density-functional tight-binding method for
  simulations of complex materials properties},\ }\href@noop {} {\bibfield
  {journal} {\bibinfo  {journal} {Phys. Rev. B}\ }\textbf {\bibinfo {volume}
  {58}},\ \bibinfo {pages} {7260} (\bibinfo {year} {1998})}\BibitemShut
  {NoStop}%
\bibitem [{\citenamefont {Mi}\ \emph {et~al.}(2016)\citenamefont {Mi},
  \citenamefont {Shao}, \citenamefont {Su}, \citenamefont {Zhou}, \citenamefont
  {Zhang}, \citenamefont {Li}, \citenamefont {Wang}, \citenamefont {Zhang},
  \citenamefont {Miao}, \citenamefont {Wang},\ and\ \citenamefont
  {Ma}}]{mi2016atlas}%
  \BibitemOpen
  \bibfield  {author} {\bibinfo {author} {\bibfnamefont {W.}~\bibnamefont
  {Mi}}, \bibinfo {author} {\bibfnamefont {X.}~\bibnamefont {Shao}}, \bibinfo
  {author} {\bibfnamefont {C.}~\bibnamefont {Su}}, \bibinfo {author}
  {\bibfnamefont {Y.}~\bibnamefont {Zhou}}, \bibinfo {author} {\bibfnamefont
  {S.}~\bibnamefont {Zhang}}, \bibinfo {author} {\bibfnamefont
  {Q.}~\bibnamefont {Li}}, \bibinfo {author} {\bibfnamefont {H.}~\bibnamefont
  {Wang}}, \bibinfo {author} {\bibfnamefont {L.}~\bibnamefont {Zhang}},
  \bibinfo {author} {\bibfnamefont {M.}~\bibnamefont {Miao}}, \bibinfo {author}
  {\bibfnamefont {Y.}~\bibnamefont {Wang}},\ and\ \bibinfo {author}
  {\bibfnamefont {Y.}~\bibnamefont {Ma}},\ }\bibfield  {title} {\bibinfo
  {title} {{ATLAS:} a real-space finite-difference implementation of
  orbital-free density functional theory},\ }\href
  {https://doi.org/https://doi.org/10.1016/j.cpc.2015.11.004} {\bibfield
  {journal} {\bibinfo  {journal} {Comput. Phys. Commun.}\ }\textbf {\bibinfo
  {volume} {200}},\ \bibinfo {pages} {87} (\bibinfo {year} {2016})}\BibitemShut
  {NoStop}%
\bibitem [{\citenamefont {Shao}\ \emph {et~al.}(2016)\citenamefont {Shao},
  \citenamefont {Mi}, \citenamefont {Xu}, \citenamefont {Wang},\ and\
  \citenamefont {Ma}}]{shao2016n}%
  \BibitemOpen
  \bibfield  {author} {\bibinfo {author} {\bibfnamefont {X.}~\bibnamefont
  {Shao}}, \bibinfo {author} {\bibfnamefont {W.}~\bibnamefont {Mi}}, \bibinfo
  {author} {\bibfnamefont {Q.}~\bibnamefont {Xu}}, \bibinfo {author}
  {\bibfnamefont {Y.}~\bibnamefont {Wang}},\ and\ \bibinfo {author}
  {\bibfnamefont {Y.}~\bibnamefont {Ma}},\ }\bibfield  {title} {\bibinfo
  {title} {O (n log n) scaling method to evaluate the ion--electron potential
  of crystalline solids},\ }\href@noop {} {\bibfield  {journal} {\bibinfo
  {journal} {J. Chem. Phys.}\ }\textbf {\bibinfo {volume} {145}} (\bibinfo
  {year} {2016})}\BibitemShut {NoStop}%
\bibitem [{\citenamefont {Thomas}(1927)}]{Thomas1927}%
  \BibitemOpen
  \bibfield  {author} {\bibinfo {author} {\bibfnamefont {L.~H.}\ \bibnamefont
  {Thomas}},\ }\bibfield  {title} {\bibinfo {title} {The calculation of atomic
  fields},\ }\href@noop {} {\bibfield  {journal} {\bibinfo  {journal} {Math.
  Proc. Cambridge Philos. Soc.}\ }\textbf {\bibinfo {volume} {23}},\ \bibinfo
  {pages} {542} (\bibinfo {year} {1927})}\BibitemShut {NoStop}%
\bibitem [{\citenamefont {Fermi}(1927)}]{Fermi1927}%
  \BibitemOpen
  \bibfield  {author} {\bibinfo {author} {\bibfnamefont {E.}~\bibnamefont
  {Fermi}},\ }\bibfield  {title} {\bibinfo {title} {Un metodo statistico per la
  determinazione di alcune priorieta dell'atome},\ }\href@noop {} {\bibfield
  {journal} {\bibinfo  {journal} {Rend. Accad. Naz. Lincei}\ }\textbf {\bibinfo
  {volume} {6}},\ \bibinfo {pages} {32} (\bibinfo {year} {1927})}\BibitemShut
  {NoStop}%
\bibitem [{\citenamefont {Fermi}(1928)}]{fermi1928statistische}%
  \BibitemOpen
  \bibfield  {author} {\bibinfo {author} {\bibfnamefont {E.}~\bibnamefont
  {Fermi}},\ }\bibfield  {title} {\bibinfo {title} {Eine statistische methode
  zur bestimmung einiger eigenschaften des atoms und ihre anwendung auf die
  theorie des periodischen systems der elemente},\ }\href@noop {} {\bibfield
  {journal} {\bibinfo  {journal} {Z. Phys.}\ }\textbf {\bibinfo {volume}
  {48}},\ \bibinfo {pages} {73} (\bibinfo {year} {1928})}\BibitemShut {NoStop}%
\bibitem [{\citenamefont {Giannozzi}\ \emph {et~al.}(2009)\citenamefont
  {Giannozzi}, \citenamefont {Baroni}, \citenamefont {Bonini}, \citenamefont
  {Calandra}, \citenamefont {Car}, \citenamefont {Cavazzoni}, \citenamefont
  {Ceresoli}, \citenamefont {Chiarotti}, \citenamefont {Cococcioni},
  \citenamefont {Dabo}, \citenamefont {Corso}, \citenamefont {de~Gironcoli},
  \citenamefont {Fabris}, \citenamefont {Fratesi}, \citenamefont {Gebauer},
  \citenamefont {Gerstmann}, \citenamefont {Gougoussis}, \citenamefont
  {Kokalj}, \citenamefont {Lazzeri}, \citenamefont {Martin-Samos},
  \citenamefont {Marzari}, \citenamefont {Mauri}, \citenamefont {Mazzarello},
  \citenamefont {Paolini}, \citenamefont {Pasquarello}, \citenamefont
  {Paulatto}, \citenamefont {Sbraccia}, \citenamefont {Scandolo}, \citenamefont
  {Sclauzero}, \citenamefont {Seitsonen}, \citenamefont {Smogunov},
  \citenamefont {Umari},\ and\ \citenamefont {Wentzcovitch}}]{quantumEspresso}%
  \BibitemOpen
  \bibfield  {author} {\bibinfo {author} {\bibfnamefont {P.}~\bibnamefont
  {Giannozzi}}, \bibinfo {author} {\bibfnamefont {S.}~\bibnamefont {Baroni}},
  \bibinfo {author} {\bibfnamefont {N.}~\bibnamefont {Bonini}}, \bibinfo
  {author} {\bibfnamefont {M.}~\bibnamefont {Calandra}}, \bibinfo {author}
  {\bibfnamefont {R.}~\bibnamefont {Car}}, \bibinfo {author} {\bibfnamefont
  {C.}~\bibnamefont {Cavazzoni}}, \bibinfo {author} {\bibfnamefont
  {D.}~\bibnamefont {Ceresoli}}, \bibinfo {author} {\bibfnamefont {G.~L.}\
  \bibnamefont {Chiarotti}}, \bibinfo {author} {\bibfnamefont {M.}~\bibnamefont
  {Cococcioni}}, \bibinfo {author} {\bibfnamefont {I.}~\bibnamefont {Dabo}},
  \bibinfo {author} {\bibfnamefont {A.~D.}\ \bibnamefont {Corso}}, \bibinfo
  {author} {\bibfnamefont {S.}~\bibnamefont {de~Gironcoli}}, \bibinfo {author}
  {\bibfnamefont {S.}~\bibnamefont {Fabris}}, \bibinfo {author} {\bibfnamefont
  {G.}~\bibnamefont {Fratesi}}, \bibinfo {author} {\bibfnamefont
  {R.}~\bibnamefont {Gebauer}}, \bibinfo {author} {\bibfnamefont
  {U.}~\bibnamefont {Gerstmann}}, \bibinfo {author} {\bibfnamefont
  {C.}~\bibnamefont {Gougoussis}}, \bibinfo {author} {\bibfnamefont
  {A.}~\bibnamefont {Kokalj}}, \bibinfo {author} {\bibfnamefont
  {M.}~\bibnamefont {Lazzeri}}, \bibinfo {author} {\bibfnamefont
  {L.}~\bibnamefont {Martin-Samos}}, \bibinfo {author} {\bibfnamefont
  {N.}~\bibnamefont {Marzari}}, \bibinfo {author} {\bibfnamefont
  {F.}~\bibnamefont {Mauri}}, \bibinfo {author} {\bibfnamefont
  {R.}~\bibnamefont {Mazzarello}}, \bibinfo {author} {\bibfnamefont
  {S.}~\bibnamefont {Paolini}}, \bibinfo {author} {\bibfnamefont
  {A.}~\bibnamefont {Pasquarello}}, \bibinfo {author} {\bibfnamefont
  {L.}~\bibnamefont {Paulatto}}, \bibinfo {author} {\bibfnamefont
  {C.}~\bibnamefont {Sbraccia}}, \bibinfo {author} {\bibfnamefont
  {S.}~\bibnamefont {Scandolo}}, \bibinfo {author} {\bibfnamefont
  {G.}~\bibnamefont {Sclauzero}}, \bibinfo {author} {\bibfnamefont {A.~P.}\
  \bibnamefont {Seitsonen}}, \bibinfo {author} {\bibfnamefont {A.}~\bibnamefont
  {Smogunov}}, \bibinfo {author} {\bibfnamefont {P.}~\bibnamefont {Umari}},\
  and\ \bibinfo {author} {\bibfnamefont {R.~M.}\ \bibnamefont {Wentzcovitch}},\
  }\bibfield  {title} {\bibinfo {title} {Quantum espresso: a modular and
  open-source software project for quantum simulations of materials},\ }\href
  {https://doi.org/10.1088/0953-8984/21/39/395502} {\bibfield  {journal}
  {\bibinfo  {journal} {J. Phys.: Condens. Matter}\ }\textbf {\bibinfo {volume}
  {21}},\ \bibinfo {pages} {395502} (\bibinfo {year} {2009})}\BibitemShut
  {NoStop}%
\bibitem [{\citenamefont {Larsen}\ \emph {et~al.}(2017)\citenamefont {Larsen},
  \citenamefont {Mortensen}, \citenamefont {Blomqvist}, \citenamefont
  {Castelli}, \citenamefont {Christensen}, \citenamefont {Dułak},
  \citenamefont {Friis}, \citenamefont {Groves}, \citenamefont {Hammer},
  \citenamefont {Hargus}, \citenamefont {Hermes}, \citenamefont {Jennings},
  \citenamefont {Jensen}, \citenamefont {Kermode}, \citenamefont {Kitchin},
  \citenamefont {Kolsbjerg}, \citenamefont {Kubal}, \citenamefont {Kaasbjerg},
  \citenamefont {Lysgaard}, \citenamefont {Maronsson}, \citenamefont {Maxson},
  \citenamefont {Olsen}, \citenamefont {Pastewka}, \citenamefont {Peterson},
  \citenamefont {Rostgaard}, \citenamefont {Schiøtz}, \citenamefont {Schütt},
  \citenamefont {Strange}, \citenamefont {Thygesen}, \citenamefont {Vegge},
  \citenamefont {Vilhelmsen}, \citenamefont {Walter}, \citenamefont {Zeng},\
  and\ \citenamefont {Jacobsen}}]{ase-paper}%
  \BibitemOpen
  \bibfield  {author} {\bibinfo {author} {\bibfnamefont {A.~H.}\ \bibnamefont
  {Larsen}}, \bibinfo {author} {\bibfnamefont {J.~J.}\ \bibnamefont
  {Mortensen}}, \bibinfo {author} {\bibfnamefont {J.}~\bibnamefont
  {Blomqvist}}, \bibinfo {author} {\bibfnamefont {I.~E.}\ \bibnamefont
  {Castelli}}, \bibinfo {author} {\bibfnamefont {R.}~\bibnamefont
  {Christensen}}, \bibinfo {author} {\bibfnamefont {M.}~\bibnamefont {Dułak}},
  \bibinfo {author} {\bibfnamefont {J.}~\bibnamefont {Friis}}, \bibinfo
  {author} {\bibfnamefont {M.~N.}\ \bibnamefont {Groves}}, \bibinfo {author}
  {\bibfnamefont {B.}~\bibnamefont {Hammer}}, \bibinfo {author} {\bibfnamefont
  {C.}~\bibnamefont {Hargus}}, \bibinfo {author} {\bibfnamefont {E.~D.}\
  \bibnamefont {Hermes}}, \bibinfo {author} {\bibfnamefont {P.~C.}\
  \bibnamefont {Jennings}}, \bibinfo {author} {\bibfnamefont {P.~B.}\
  \bibnamefont {Jensen}}, \bibinfo {author} {\bibfnamefont {J.}~\bibnamefont
  {Kermode}}, \bibinfo {author} {\bibfnamefont {J.~R.}\ \bibnamefont
  {Kitchin}}, \bibinfo {author} {\bibfnamefont {E.~L.}\ \bibnamefont
  {Kolsbjerg}}, \bibinfo {author} {\bibfnamefont {J.}~\bibnamefont {Kubal}},
  \bibinfo {author} {\bibfnamefont {K.}~\bibnamefont {Kaasbjerg}}, \bibinfo
  {author} {\bibfnamefont {S.}~\bibnamefont {Lysgaard}}, \bibinfo {author}
  {\bibfnamefont {J.~B.}\ \bibnamefont {Maronsson}}, \bibinfo {author}
  {\bibfnamefont {T.}~\bibnamefont {Maxson}}, \bibinfo {author} {\bibfnamefont
  {T.}~\bibnamefont {Olsen}}, \bibinfo {author} {\bibfnamefont
  {L.}~\bibnamefont {Pastewka}}, \bibinfo {author} {\bibfnamefont
  {A.}~\bibnamefont {Peterson}}, \bibinfo {author} {\bibfnamefont
  {C.}~\bibnamefont {Rostgaard}}, \bibinfo {author} {\bibfnamefont
  {J.}~\bibnamefont {Schiøtz}}, \bibinfo {author} {\bibfnamefont
  {O.}~\bibnamefont {Schütt}}, \bibinfo {author} {\bibfnamefont
  {M.}~\bibnamefont {Strange}}, \bibinfo {author} {\bibfnamefont {K.~S.}\
  \bibnamefont {Thygesen}}, \bibinfo {author} {\bibfnamefont {T.}~\bibnamefont
  {Vegge}}, \bibinfo {author} {\bibfnamefont {L.}~\bibnamefont {Vilhelmsen}},
  \bibinfo {author} {\bibfnamefont {M.}~\bibnamefont {Walter}}, \bibinfo
  {author} {\bibfnamefont {Z.}~\bibnamefont {Zeng}},\ and\ \bibinfo {author}
  {\bibfnamefont {K.~W.}\ \bibnamefont {Jacobsen}},\ }\bibfield  {title}
  {\bibinfo {title} {The atomic simulation environment—a python library for
  working with atoms},\ }\href
  {http://stacks.iop.org/0953-8984/29/i=27/a=273002} {\bibfield  {journal}
  {\bibinfo  {journal} {J. Phys.: Condens. Matter}\ }\textbf {\bibinfo {volume}
  {29}},\ \bibinfo {pages} {273002} (\bibinfo {year} {2017})}\BibitemShut
  {NoStop}%
\bibitem [{\citenamefont {Huang}\ and\ \citenamefont
  {Carter}(2008)}]{huang2008transferable}%
  \BibitemOpen
  \bibfield  {author} {\bibinfo {author} {\bibfnamefont {C.}~\bibnamefont
  {Huang}}\ and\ \bibinfo {author} {\bibfnamefont {E.~A.}\ \bibnamefont
  {Carter}},\ }\bibfield  {title} {\bibinfo {title} {Transferable local
  pseudopotentials for magnesium, aluminum and silicon},\ }\href@noop {}
  {\bibfield  {journal} {\bibinfo  {journal} {Phys. Chem. Chem. Phys.}\
  }\textbf {\bibinfo {volume} {10}},\ \bibinfo {pages} {7109} (\bibinfo {year}
  {2008})}\BibitemShut {NoStop}%
\bibitem [{\citenamefont {Perdew}\ and\ \citenamefont
  {Zunger}(1981)}]{Perdew1981PZ}%
  \BibitemOpen
  \bibfield  {author} {\bibinfo {author} {\bibfnamefont {J.~P.}\ \bibnamefont
  {Perdew}}\ and\ \bibinfo {author} {\bibfnamefont {A.}~\bibnamefont
  {Zunger}},\ }\bibfield  {title} {\bibinfo {title} {Self-interaction
  correction to density-functional approximations for many-electron systems},\
  }\href {https://doi.org/10.1103/PhysRevB.23.5048} {\bibfield  {journal}
  {\bibinfo  {journal} {Phys. Rev. B}\ }\textbf {\bibinfo {volume} {23}},\
  \bibinfo {pages} {5048} (\bibinfo {year} {1981})}\BibitemShut {NoStop}%
\bibitem [{\citenamefont {Boyer}\ \emph {et~al.}(1991)\citenamefont {Boyer},
  \citenamefont {Kaxiras}, \citenamefont {Feldman}, \citenamefont {Broughton},\
  and\ \citenamefont {Mehl}}]{boyer1991new}%
  \BibitemOpen
  \bibfield  {author} {\bibinfo {author} {\bibfnamefont {L.}~\bibnamefont
  {Boyer}}, \bibinfo {author} {\bibfnamefont {E.}~\bibnamefont {Kaxiras}},
  \bibinfo {author} {\bibfnamefont {J.}~\bibnamefont {Feldman}}, \bibinfo
  {author} {\bibfnamefont {J.}~\bibnamefont {Broughton}},\ and\ \bibinfo
  {author} {\bibfnamefont {M.}~\bibnamefont {Mehl}},\ }\bibfield  {title}
  {\bibinfo {title} {New low-energy crystal structure for silicon},\
  }\href@noop {} {\bibfield  {journal} {\bibinfo  {journal} {Phys. Rev. Lett.}\
  }\textbf {\bibinfo {volume} {67}},\ \bibinfo {pages} {715} (\bibinfo {year}
  {1991})}\BibitemShut {NoStop}%
\bibitem [{\citenamefont {Yin}\ and\ \citenamefont
  {Cohen}(1982)}]{yin1982theory}%
  \BibitemOpen
  \bibfield  {author} {\bibinfo {author} {\bibfnamefont {M.}~\bibnamefont
  {Yin}}\ and\ \bibinfo {author} {\bibfnamefont {M.~L.}\ \bibnamefont
  {Cohen}},\ }\bibfield  {title} {\bibinfo {title} {Theory of static structural
  properties, crystal stability, and phase transformations: Application to si
  and ge},\ }\href@noop {} {\bibfield  {journal} {\bibinfo  {journal} {Phys.
  Rev. B}\ }\textbf {\bibinfo {volume} {26}},\ \bibinfo {pages} {5668}
  (\bibinfo {year} {1982})}\BibitemShut {NoStop}%
\bibitem [{\citenamefont {Wang}\ \emph {et~al.}(2010)\citenamefont {Wang},
  \citenamefont {Lv}, \citenamefont {Zhu},\ and\ \citenamefont
  {Ma}}]{calypso1}%
  \BibitemOpen
  \bibfield  {author} {\bibinfo {author} {\bibfnamefont {Y.}~\bibnamefont
  {Wang}}, \bibinfo {author} {\bibfnamefont {J.}~\bibnamefont {Lv}}, \bibinfo
  {author} {\bibfnamefont {L.}~\bibnamefont {Zhu}},\ and\ \bibinfo {author}
  {\bibfnamefont {Y.}~\bibnamefont {Ma}},\ }\bibfield  {title} {\bibinfo
  {title} {Crystal structure prediction via particle-swarm optimization},\
  }\href {https://doi.org/10.1103/PhysRevB.82.094116} {\bibfield  {journal}
  {\bibinfo  {journal} {Phys. Rev. B}\ }\textbf {\bibinfo {volume} {82}},\
  \bibinfo {pages} {094116} (\bibinfo {year} {2010})}\BibitemShut {NoStop}%
\bibitem [{\citenamefont {Wang}\ \emph {et~al.}(2012)\citenamefont {Wang},
  \citenamefont {Lv}, \citenamefont {Zhu},\ and\ \citenamefont
  {Ma}}]{calypso2}%
  \BibitemOpen
  \bibfield  {author} {\bibinfo {author} {\bibfnamefont {Y.}~\bibnamefont
  {Wang}}, \bibinfo {author} {\bibfnamefont {J.}~\bibnamefont {Lv}}, \bibinfo
  {author} {\bibfnamefont {L.}~\bibnamefont {Zhu}},\ and\ \bibinfo {author}
  {\bibfnamefont {Y.}~\bibnamefont {Ma}},\ }\bibfield  {title} {\bibinfo
  {title} {Calypso: A method for crystal structure prediction},\ }\href
  {https://doi.org/https://doi.org/10.1016/j.cpc.2012.05.008} {\bibfield
  {journal} {\bibinfo  {journal} {Comput. Phys. Commun.}\ }\textbf {\bibinfo
  {volume} {183}},\ \bibinfo {pages} {2063} (\bibinfo {year}
  {2012})}\BibitemShut {NoStop}%
\bibitem [{sup()}]{supp}%
  \BibitemOpen
  \href@noop {} {}\bibinfo {note} {See Supplemental Material at
  URL-will-be-inserted-by-publisher for the data.}\BibitemShut {Stop}%
\bibitem [{\citenamefont {Murnaghan}(1944)}]{Murnaghan1944Compressibility}%
  \BibitemOpen
  \bibfield  {author} {\bibinfo {author} {\bibfnamefont {F.}~\bibnamefont
  {Murnaghan}},\ }\bibfield  {title} {\bibinfo {title} {The compressibility of
  media under extreme pressures},\ }\href@noop {} {\bibfield  {journal}
  {\bibinfo  {journal} {Proc. Natl. Acad. Sci. USA}\ }\textbf {\bibinfo
  {volume} {30}},\ \bibinfo {pages} {244} (\bibinfo {year} {1944})}\BibitemShut
  {NoStop}%
\end{thebibliography}%
\end{document}